\documentclass{article} % [draft]: no print figs

\usepackage{arxiv-jrr}

\usepackage[utf8]{inputenc} % allow utf-8 input
\usepackage[T1]{fontenc}    % use 8-bit T1 fonts
\usepackage{hyperref}       % hyperlinks
\usepackage{url}            % simple URL typesetting
\usepackage{booktabs}       % professional-quality tables
\usepackage{amsfonts}       % blackboard math symbols
\usepackage{nicefrac}       % compact symbols for 1/2, etc.
\usepackage{microtype}      % microtypography
\usepackage{lipsum}		% Can be removed after putting your text content
\usepackage{graphicx}
\usepackage{natbib}
\usepackage{doi}
\usepackage{amsmath}

\title{Automating the Expert Eye: \\A System-Agnostic Deep Learning Framework for Rare Event Discovery in Imbalanced Force Spectroscopy
}

\author{ \href{https://orcid.org/0000-0001-9837-6257}{\includegraphics[scale=0.06]{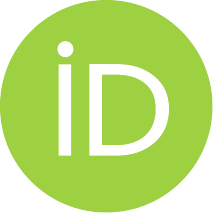}\hspace{1mm}Jorge~Rodriguez-Ramos} \\
	Independent Researcher\\
	Palermo, Italy\\
	\\
	\texttt{pro@jrr.one} \\
}

\renewcommand{\shorttitle}{Automating the Expert Eye in SMFS}

\hypersetup{
pdftitle={Automating the Expert Eye: A System-Agnostic Deep Learning Framework for Rare Event Discovery in Imbalanced Force Spectroscopy},
pdfsubject={cs.AI},
pdfauthor={Jorge Rodriguez-Ramos},
pdfkeywords={Single-molecule force spectroscopy, Atomic force microscopy, Deep learning, Transfer learning, Class imbalance, Model interpretability},
}

\begin{document}
\maketitle

\begin{abstract}
	Single-molecule force spectroscopy (SMFS) probes biomolecular mechanics at the single-molecule level, but high-throughput acquisition of force-extension curves creates a data processing bottleneck. Identifying rare specific unbinding events among thousands of noise-dominated traces still relies largely on manual visual inspection. Here, we present a deep learning workflow for the preliminary screening of severely imbalanced SMFS datasets. Using 2D image representations of force-extension curves without system-specific physical priors, we fine-tuned a modified ResNet18 architecture trained with focal loss and evaluated it on the mechanical unfolding pathways of the \textit{Ruminococcus champanellensis} XMod-Doc/Coh complex. On a held-out test set with a 1.34\% target prevalence (13 positive $P_2$ events among 970 curves), the model achieved 92.31\% recall (12 of 13 events) and 91.95\% specificity (880 of 957 negative traces), automatically discarding 90.82\% of the dataset and reducing manual review from 970 to 89 flagged curves. We also integrated Gradient-weighted Class Activation Mapping (Grad-CAM) to provide a qualitative visual reference of the curve regions driving each prediction. The complete Python pipeline is released as open-source software with an executable Google Colab notebook, enabling reproducible SMFS screening without dedicated local GPU hardware.
\end{abstract}


\section{Introduction}

Single-molecule force spectroscopy (SMFS)
based on atomic force microscopy (AFM)
enables the mechanical manipulation of individual biomolecular complexes
under physiological conditions \citep{neuman2008single,rief1997reversible}.
This approach provides quantitative access to cellular adhesion,
protein unfolding dynamics, and receptor-ligand energy landscapes.

Despite its high force resolution, SMFS faces a major data-filtering bottleneck. Modern high-throughput SMFS protocols routinely generate thousands of raw force-extension trajectories to achieve statistical significance for rare molecular interactions. However, due to stochastic single-molecule tethering and binding, typically 90\%--99\% of these traces consist of non-specific surface adhesion, instrumentation artifacts, cantilever ringing, or baseline noise. Extracting specific unbinding events from such large datasets remains a labor-intensive, manual process. As high-speed and multiplexed force spectroscopy techniques continue to increase data acquisition rates \citep{rico2013high}, reliance on visual inspection by experienced operators fails to scale and introduces selection bias.

Early automated screening methods relied on deterministic filtering, such as multi-peak fitting with the worm-like chain (WLC) model \citep{marko1995stretching}. Although effective for isolated, single-domain proteins, WLC fitting routinely breaks down when traces contain overlapping non-specific adhesion or multi-state unfolding pathways. This limitation has motivated the development of standardized, open-source analysis frameworks for complex force-spectroscopy datasets \citep{lopez2024pyfmlab}.

An alternative approach bypasses rigid curve-fitting by changing the data representation \citep{doffini2023iterative}. By rasterizing 1D force-extension trajectories into square 2D images, the authors treated force curves as visual patterns, enabling deep convolutional neural networks (CNNs) to automate the visual inspection typically performed by experienced operators. Their FUSION framework couples a DenseNet121 architecture with a triplet-loss embedding layer to embed and cluster distinct single-molecule unfolding pathways.

Although 2D image representations make SMFS traces amenable to deep learning, two practical obstacles limit routine laboratory adoption:

\begin{itemize}
	\item \textbf{Severe class imbalance:} Target conformations, rare intermediates, or low-yield mutant pathways often comprise less than 2\% of raw traces. Under standard cross-entropy loss, the majority background class dominates training, producing high false-negative rates that miss the rare events of interest.

	\item \textbf{Lack of interpretability:} Standard neural networks provide no indication of which trace features govern their classifications. Without transparent feature attribution, researchers remain reluctant to let automated pipelines discard most of an experimental dataset.
\end{itemize}

We address these limitations with an end-to-end, system-agnostic computer vision pipeline for automated SMFS trace screening and rare-event detection under severe class imbalance. Adopting the existing 2D image representation \citep{doffini2023iterative} rather than introducing a new transformation isolates the effect of our architectural modifications for a direct benchmark.

Our pipeline combines three components within a lightweight architecture:

\begin{enumerate}
	\item \textbf{Focal-loss classification:} A modified ResNet18 trained with focal loss \citep{lin2017focal} down-weights easily classified background traces, concentrating gradient updates on rare molecular rupture signatures.
	\item \textbf{Feature attribution via Grad-CAM:} Gradient-weighted Class Activation Mapping (Grad-CAM) \citep{selvaraju2017grad} projects class activation maps onto classified traces, allowing operators to inspect whether predictions rely on physical features (such as WLC stretching regions) rather than instrumentation noise.
	\item \textbf{Dual-threshold filtering:} An adjustable dual-threshold gate replaces a rigid binary cutoff, discarding obvious noise, retaining high-confidence events, and flagging borderline traces for manual inspection.
\end{enumerate}

We benchmark this workflow under severe class imbalance using the public single-molecule dataset of the multi-domain XModule-Dockerin/Cohesin (XMod-Doc/Coh) complex from \textit{Ruminococcus champanellensis} \citep{doffini2023iterative}, whose multiple unbinding pathways produce force-extension profiles that are difficult to distinguish visually. On a test set with a 1.34\% target class prevalence, the pipeline isolates these rare unbinding events with high recall while automatically discarding over 90\% of background traces. The workflow is released as an open-source tool compatible with standard cloud-based GPU environments, supporting automated SMFS trace screening without specialized local hardware.

\section{Materials and Methods}
\label{sec:methods}

\subsection{Dataset Origin and Biomolecular System}
We evaluated our pipeline on the public SMFS dataset of the \textit{Ruminococcus champanellensis} XMod-Doc/Coh multi-domain complex \citep{doffini2023iterative}, allowing direct comparison against an established benchmark.

The dataset contains three primary unbinding pathways:
\begin{itemize}
	\item \textbf{Pathway 1 ($P_1$):} Direct, high-force complex rupture without prior unfolding of the adjacent mechanostable Ig-like XMod domain.
	\item \textbf{Pathway 2 ($P_2$):} Sequential unfolding in which the XMod domain unfolds prior to complex dissociation, yielding a characteristic contour length increment ($\Delta L_c \approx 29\text{--}32 \text{ nm}$).
	\item \textbf{Pathway 3 ($P_3$):} Low-force unbinding with a profile similar to $P_1$ but a substantially lower rupture force, making it difficult to distinguish from baseline noise.
\end{itemize}

\subsection{2D Image Representation of Force Curves}

We used the pre-rasterized 2D experimental dataset from \citet{doffini2023iterative} to ensure direct comparability and preserve the force-extension features, including WLC entropic elasticity curves and sharp vertical rupture drops. Rather than regenerating the image transformations locally, we downloaded the dataset directly from the public Zenodo repository:

\begin{center}
	\url{https://zenodo.org/records/8224237}
\end{center}

In this format, normalized extension ($x$-axis) and force ($y$-axis) coordinates are mapped onto $200 \times 200$ pixel arrays, which we loaded directly into the training pipeline as NumPy arrays (\texttt{x\_2D\_processed.npy} for the 2D images and \texttt{y\_raw.npy} for the interaction labels). Using these preprocessed arrays avoids intermediate image-file conversions and ensures that the inputs to our convolutional backbone remain identical to the benchmark dataset without additional preprocessing beyond resizing to $224 \times 224$ pixels for ResNet18.

\subsection{Dataset Partitioning and Stratification}

The dataset comprises 1495 experimental force curves. We formulated a binary classification task with Pathway 2 ($P_2$, $n = 128$; 8.56\%) as the positive target class, while grouping the remaining pathways ($P_1$, $n = 254$; $P_3$, $n = 91$) and non-specific or noise traces ($n = 1022$) into the negative class ($n = 1367$).

We partitioned the dataset into independent training, validation, and held-out test subsets to prevent data leakage. First, we constructed a severely imbalanced test set with a target prevalence of 1.34\% to evaluate model performance under realistic rare-event conditions. We then applied an 80/20 stratified split to the remaining 525 curves (21.9\% positive) to preserve class proportions between training and validation:

\begin{itemize}
	\item \textbf{Training set:} 420 trajectories (328 negative, 92 positive), used for model parameter optimization.
	\item \textbf{Validation set:} 105 trajectories (82 negative, 23 positive), used for hyperparameter tuning, focal-loss weighting, and dual-threshold calibration.
	\item \textbf{Test set:} 970 held-out trajectories (957 negative, 13 positive; 1.34\% positive prevalence), used exclusively for final evaluation under severe class imbalance.
\end{itemize}

\subsection{Model Architecture and Training}

\begin{figure}[htbp]
	\centering
	\includegraphics[width=\textwidth]{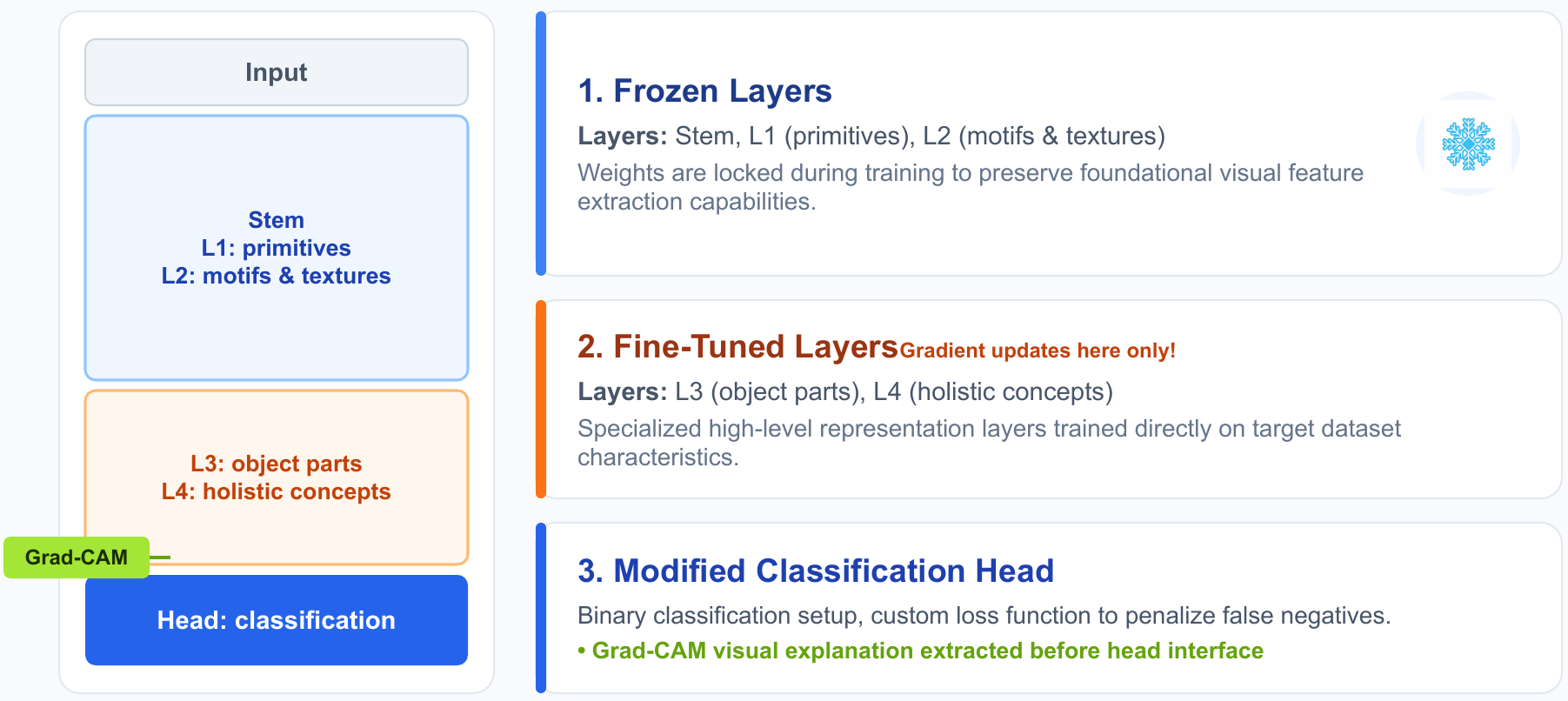}
	\caption{Schematic of the modified ResNet18 architecture.}
	\label{fig:resnet}
\end{figure}

We implemented a modified ResNet18 architecture (Figure~\ref{fig:resnet}) to enable high-throughput classification on standard cloud-based GPUs (e.g., Google Colab) without requiring dedicated local hardware. Using transfer learning from ImageNet pretraining, we froze the early convolutional blocks (\texttt{layer1} and \texttt{layer2}), which capture general low-level features such as edges, lines, corners, and curves, and fine-tuned only the final two blocks (\texttt{layer3} and \texttt{layer4}) to learn higher-level features characteristic of SMFS force-extension profiles. We replaced the original 1000-class output layer with a custom binary classification head comprising a 256-node linear layer, a ReLU activation, a 40\% dropout layer to reduce overfitting, and a two-node linear output corresponding to the negative and positive classes.

We replaced standard cross-entropy with focal loss \citep{lin2017focal} to address class imbalance during training and improve sensitivity to rare events:
\[
\mathrm{FL}(p_t) = -\alpha_t (1 - p_t)^\gamma \log(p_t)
\]
where $p_t$ is the predicted probability for the true class. Setting the focusing parameter to $\gamma = 2.0$ down-weights the loss contribution of easily classified background noise. In addition, class weights ($\alpha = [0.3, 0.7]$ for the negative and positive classes, respectively) increase the penalty for false negatives, concentrating training updates on the sparse features of specific molecular ruptures.

\subsection{Model Interpretability via Grad-CAM}
We used Gradient-weighted Class Activation Mapping (Grad-CAM) \citep{selvaraju2017grad} to visualize which trace regions drive classification. Grad-CAM computes the gradient of the target class score $c$ with respect to the feature maps $A^k$ of the final convolutional block (\texttt{layer4}), applies global average pooling to obtain channel weights $\alpha_k^c$, and combines them into a class activation map:
\[
L_{\text{Grad-CAM}}^c = \text{ReLU}\left(\sum_k \alpha_k^c A^k\right)
\]
The activation maps were upsampled and overlaid onto the $224 \times 224$ pixel ResNet18 inputs (resized from the original $200 \times 200$ NumPy arrays), enabling visual inspection of whether predictions align with physical features (such as WLC stretching regions) rather than image artifacts.

\subsection{Computational Hardware and Implementation}
All training and evaluation runs were executed locally on a standard workstation equipped with an Intel Core Ultra 5 CPU, 32 GB RAM, and an NVIDIA RTX 2000 Ada Generation GPU (16 GB VRAM). Because the workflow fits well within a 16 GB VRAM footprint, it is also compatible with standard cloud GPU instances such as the NVIDIA T4 available on Google Colab. The pipeline was implemented in Python (v3.10) using PyTorch (v2.7.1) and NumPy; the complete source code and a runnable Google Colab notebook are publicly available (see Code Availability).

\section{Results and Discussion}

\subsection{Model Optimization}
We trained the modified ResNet18 network for 100 epochs on the training set (420 trajectories) and evaluated performance after each epoch on the independent validation set (105 trajectories) to monitor overfitting and select the final checkpoint. Total training time was 737~s (12.28~min) on the local workstation.

\begin{figure}[htbp]
	\centering
	\includegraphics[width=\textwidth]{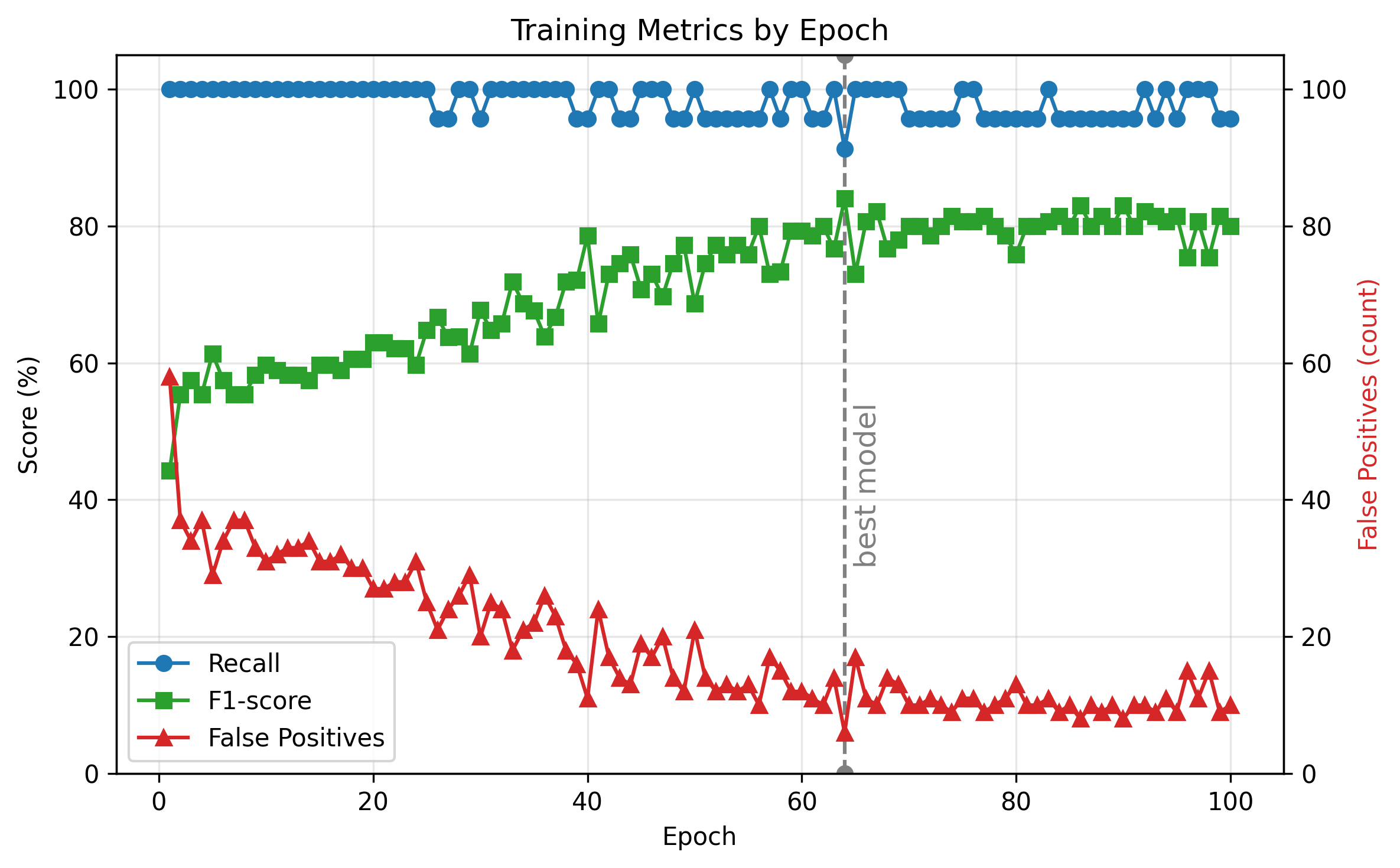}
	\caption{Validation performance during 100 epochs of training for the modified ResNet18 model. Recall (blue) remained high throughout training, whereas early epochs produced a high number of false positives (red). Training with focal loss steadily improved the validation F1-score (green), reaching a peak of 84.00\% at epoch 64 (dashed line), which was selected as the final model checkpoint.}
	\label{fig:training_curve}
\end{figure}

Because the filtering pipeline must balance the recovery of rare $P_2$ events (recall) against the rejection of background noise (precision), we used the validation F1-score as our primary selection metric (Figure~\ref{fig:training_curve}). During the early epochs, the model achieved 100\% recall at the expense of a high false-positive rate. As training progressed, focal loss reduced false positives on borderline traces. Validation F1-score reached its peak of 84.00\% at epoch 64, after which performance declined slightly and stabilized; we selected the weights from epoch 64 for all subsequent evaluations.

Standard binary classification uses a single probability threshold of $p = 0.50$, classifying traces above this value as positive and those below as negative. To calibrate our dual-threshold filter, we evaluated the epoch 64 model on the validation set across a range of probability cutoffs (Table~\ref{tab:threshold_calibration}). Lowering the threshold to $p = 0.40$ reduced false negatives from 2 to 1 but more than doubled false positives (from 6 to 13). Conversely, raising the threshold to $p = 0.80$ eliminated false positives entirely on the validation set, at the cost of missing 14 of the 23 positive traces.

Based on this trade-off, we set the lower and upper reference thresholds to $p = 0.50$ and $p = 0.80$, respectively, for screening unseen traces: curves with $p \le 0.50$ are discarded as negative, curves in the intermediate range ($0.50 < p \le 0.80$) are flagged for manual review, and curves with $p > 0.80$ are automatically accepted as positive.

\begin{table}[htbp]
	\centering
	\caption{Validation-set false positives and false negatives across probability thresholds for the epoch 64 model ($n = 105$; 82 negative, 23 positive). Selected lower and upper dual-threshold boundaries are highlighted in bold.}
	\label{tab:threshold_calibration}
	\begin{tabular}{@{}lcc@{}}
		\toprule
		\textbf{Threshold ($p$)}      & \textbf{False Positives} & \textbf{False Negatives} \\ \midrule
		0.30                          &            21            &            0             \\
		0.40                          &            13            &            1             \\
		\textbf{0.50 (Low Boundary)}  &       \textbf{6}        &        \textbf{2}        \\
		0.60                          &            4             &            3             \\
		0.70                          &            3             &            5             \\
		\textbf{0.80 (High Boundary)} &        \textbf{0}        &       \textbf{14}        \\
		0.85                          &            0             &            20            \\
		0.90                          &            0             &            21            \\
		0.95                          &            0             &            23            \\
		0.98                          &            0             &            23            \\ \bottomrule
	\end{tabular}
\end{table}

\subsection{Classification Performance on the Test Dataset}
We evaluated the selected epoch 64 model on the held-out test set of 970 force curves, in which the target $P_2$ pathway accounted for 1.34\% of the data (13 positive events and 957 negative traces).

At the lower threshold of $p = 0.50$, the model achieved a recall (true positive rate) of 92.31\%, identifying 12 of the 13 positive $P_2$ events with only 1 false negative (Table~\ref{tab:confusion_matrix}). It also correctly rejected 880 of the 957 negative traces, corresponding to a specificity of 91.95\%. In total, 881 curves (90.82\% of the test set) were automatically discarded below the lower threshold.

Conversely, 77 negative traces were misclassified as false positives (precision = 13.48\%), reflecting the expected base-rate effect at 1.34\% target prevalence. Even without applying the upper threshold, this filtering step reduced the manual screening workload by an order of magnitude, from 970 curves down to 89 flagged candidates.

\begin{table}[htbp]
	\centering
	\caption{Confusion matrix for the epoch 64 model on the held-out test set ($n = 970$) at the lower threshold ($p = 0.50$). Rows indicate the ground-truth class and columns indicate the model prediction.}
	\label{tab:confusion_matrix}
	\begin{tabular}{@{}lcc@{}}
		\toprule
		& \multicolumn{2}{c}{\textbf{Predicted Class}} \\ \cmidrule(l){2-3}
		\textbf{True Class} & \textbf{Positive ($P_2$)} & \textbf{Negative} \\ \midrule
		\textbf{Positive ($P_2$)} & 12 & 1 \\
		\textbf{Negative} & 77 & 880 \\ \bottomrule
	\end{tabular}
\end{table}

\subsection{Grad-CAM Visual Inspection}

\begin{figure}[htbp]
	\centering
	\includegraphics[width=.8\textwidth]{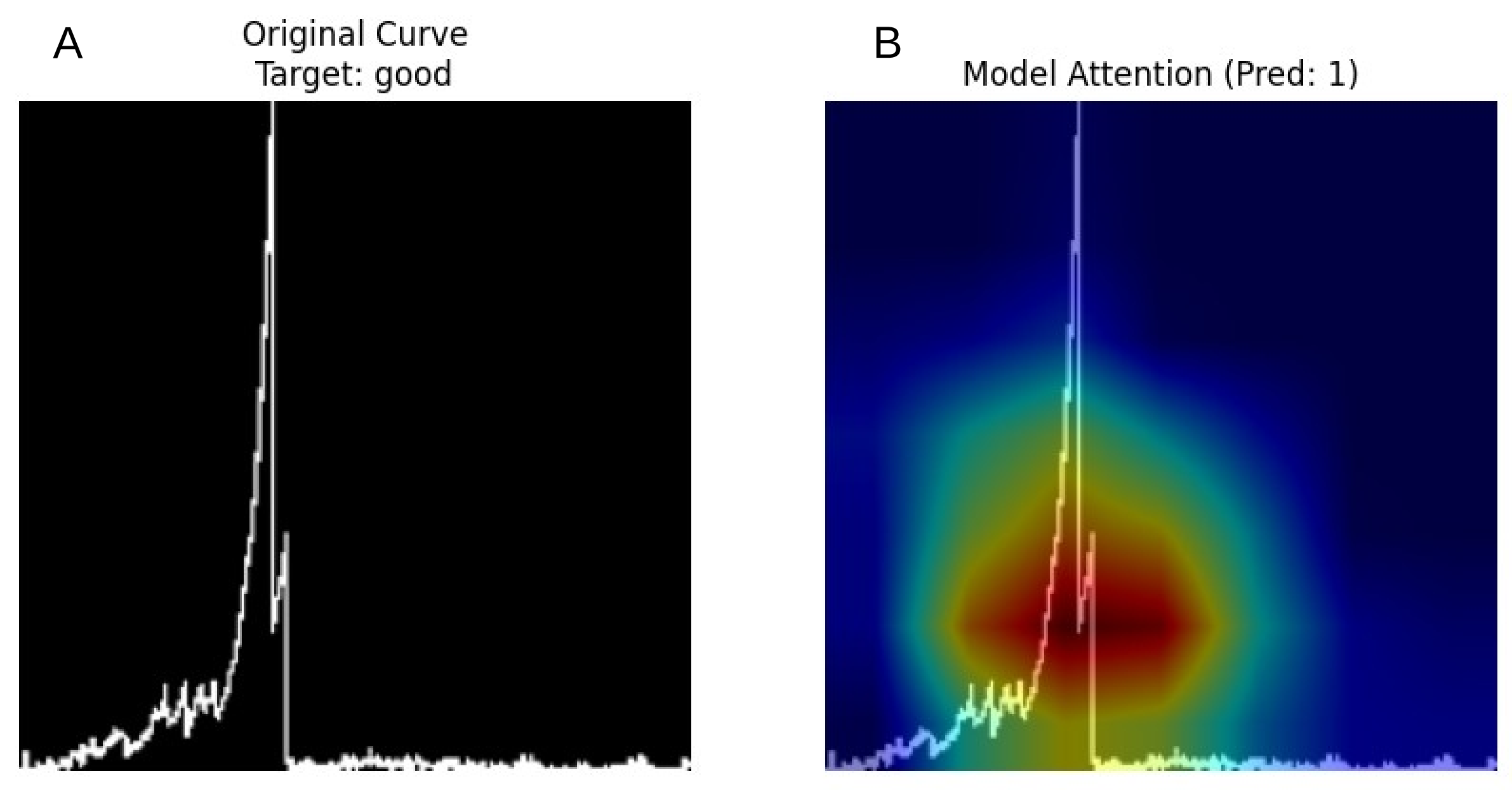}
	\caption{Qualitative visual reference using Gradient-weighted Class Activation	 Mapping (Grad-CAM). \textbf{(A)} Original 1D force-extension trajectory of a true positive $P_2$ unbinding event. \textbf{(B)} Rasterized trajectory overlaid with the upsampled class activation map extracted from \texttt{layer4}. The model classified the trace as positive (Pred: 1, $p > 0.50$), with the highest activations localized over the stretching and rupture region rather than the background.}
	\label{fig:gradcam}
\end{figure}

The ``black-box'' nature of deep neural networks often limits trust in automated SMFS screening. To provide a qualitative visual reference alongside the predicted probability, we overlaid Grad-CAM class activation maps from \texttt{layer4} onto the classified traces (Figure~\ref{fig:gradcam}).

For a representative true positive $P_2$ trajectory (Figure~\ref{fig:gradcam}B), the highest activations coincide with the mechanically relevant region of the force-extension profile rather than empty background pixels or image artifacts. Because Grad-CAM is a post-hoc interpretability tool, these maps do not constitute proof of a specific molecular unfolding event; instead, they provide an immediate sanity check that helps operators spot spurious predictions during manual review of intermediate-confidence force curves.

\section{Conclusion}
We demonstrated that combining 2D force-curve representations with a fine-tuned ResNet18 architecture and focal loss enables automated screening of SMFS datasets under severe class imbalance. On a held-out test set with a 1.34\% target prevalence of $P_2$ unbinding events, the model achieved 92.31\% recall (12 of 13 positive events) and 91.95\% specificity (880 of 957 negative traces), automatically discarding 90.82\% of the dataset and reducing the manual review workload from 970 to 89 force curves.

Because the classifier operates directly on 2D force-extension images without hard-coded physical priors (such as system-specific contour lengths or rupture-force thresholds), the workflow is system-agnostic by design: although benchmarked here on the XMod-Doc/Coh complex, it can be adapted to other SMFS systems by fine-tuning the network on a labeled subset of the target experiment.

Pairing the classifier with Grad-CAM class activation maps provides a qualitative visual reference alongside the predicted probability, allowing operators to check whether predictions align with mechanically relevant regions of the force-extension profile during manual review of flagged curves. The complete Python workflow is released as open-source software with an executable Google Colab notebook, enabling laboratories to adapt and run automated SMFS screening without dedicated local GPU hardware.

\section*{Code Availability}
The complete source code, training and evaluation scripts, and an executable Google Colab notebook for reproducing the workflow are publicly available on GitHub at \url{https://github.com/jorge-rodriguez-ramos/afm_jrr}.

\section*{Acknowledgements}
The author thanks F. Rico for fruitful discussions that helped shape the intellectual direction of this work. Additionally, the author acknowledges the use of Google Gemini (Google LLC) strictly as a conversational assistant to refine sentence structure and format \LaTeX\ code during the preparation of this manuscript. The author independently generated all data, conducted the primary analysis, and takes full responsibility for the scientific accuracy and final content of the publication.

\bibliographystyle{unsrtnat}
\bibliography{references}  %%% Uncomment this line and comment out the ``thebibliography'' section below to use the external .bib file (using bibtex) .

\end{document}